\documentclass[prd,aps,twocolumn,nofootinbib]{revtex4}

\usepackage[utf8]{inputenc}

\usepackage{graphicx}
\usepackage{color}
\usepackage{hyperref}
\usepackage{xcolor}
\usepackage{amsmath}
\usepackage{multirow}
\usepackage{array}
\usepackage{subcaption}
\hypersetup{
    colorlinks,
    linkcolor={red!80!black},
    citecolor={blue!80!black},
    urlcolor={green!50!black}
}

\def\aap{Astron.\ Astrophys.}
\def\apj{Astropys.\ J.}

\def\physrep{Phys.\ Rep.}

\def\d{{\mathrm d}}

\def\e{{\mathrm e}}

\def\theta{\vartheta}

\def\be{\begin{equation}}
\def\ee{\end{equation}}
\def\ba{\begin{eqnarray}}
\def\ea{\end{eqnarray}}
\def\nn{\nonumber}

\def\lsim{\raise0.3ex\hbox{$\;<$\kern-0.75em\raise-1.1ex\hbox{$\sim\;$}}}
\def\gsim{\raise0.3ex\hbox{$\;>$\kern-0.75em\raise-1.1ex\hbox{$\sim\;$}}}

\begin{document}

\title{Energy dependence of the knee in the cosmic-ray spectrum across the Milky Way}
\author{C.~Prevotat$^{1}$, M.~Kachelrie\ss$^{2}$, S.~Koldobskiy$^{3}$, A.~Neronov$^{1,4}$  and D.~Semikoz$^{1}$}
\affiliation{$^1$Université de Paris Cite, CNRS, Astroparticule et Cosmologie, F-75013 Paris, France}
\affiliation{$^2$Institutt for fysikk, NTNU, Trondheim, Norway}
\affiliation{$^{3}$Sodankyl\"a Geophysical Observatory and Space Physics and Astronomy Research Unit, University of Oulu, 90014 Oulu, Finland}
\affiliation{$^4$Laboratory of Astrophysics, Ecole Polytechnique Federale de Lausanne, CH-1015, Lausanne, Switzerland}
\begin{abstract}
The all-particle spectrum of cosmic rays measured at Earth has a knee-like 
feature around 4\,PeV. A priori, it is not clear if this is a local feature 
specific to the Solar neighbourhood in the Milky Way, or if it is a generic 
property of the Galactic cosmic-ray spectrum. We argue that combining gamma-ray 
and cosmic-ray data of  LHAASO indicates that the knee is a local feature. In order
to demonstrate this, we derive a model for the local cosmic-ray spectrum and composition, 
consistent with the recent LHAASO measurements of the all-particle spectrum and the mean 
logarithmic mass in the knee region. We calculate the spectrum of diffuse gamma-ray emission 
based on this model and find that the expected spectral shape of the diffuse gamma-ray flux 
disagrees with the LHAASO measurements of the diffuse gamma-ray emission in 
the 10-100\,TeV energy range in the inner and outer Galaxy. We determine the 
break energy in the CR spectrum expected from these gamma-ray data and find it an energy 
ten times lower than obtained from local measurements.
\end{abstract}
\maketitle

\section{Introduction}

The all-particle energy spectrum of high-energy cosmic rays (CRs) at Earth has been measured with increasing precision over the last sixty
years. At the energy $E_{\rm k}\simeq 4$\,PeV, the most pronounced change
dubbed the CR knee occurs where the spectral index $\d N/\d E\propto E^{-\beta}$
changes from $\beta\simeq 2.7$ below to $\beta\simeq 3.1$ above the knee. While there
is a general agreement between various experiments about the position of the
knee in the all-particle spectrum, there have been substantial differences concerning
the elemental composition of the CR flux in the knee region measured by different 
experiments~\cite{Kachelriess:2019oqu}.  The LHAASO experiment has the potential to 
improve the measurements in this energy range considerably, with the recent precise
measurement of the mean logarithmic mass as a first step in this 
direction~\cite{data_LHAASO:2024}.

Explanations for the origin of the knee fall in two main classes of 
models~\cite{Kachelriess:2019oqu}. In the
first one, the knee is caused by a change in the propagation of
Galactic CRs around the PeV energy, where a change in 
the energy dependence of the confinement time induces a steepening of the CR 
spectrum~\cite{1971CoASP...3..155S,1993A&A...268..726P,Candia:2002we,Candia:2003dk,Giacinti:2014xya,Giacinti:2015hva}.
Alternatively, the knee may be connected to properties of the
injection spectrum of Galactic CRs. For instance, the knee might correspond
to a break in the source CR energy spectrum~\cite{Drury:2003fd,Cardillo:2015zda}
or to the maximal rigidity to which the CR source population dominating the CR flux 
below PeV can accelerate~\cite{Stanev:1993tx,Kobayakawa:2000nq,Hillas:2005cs}.
A variation of this model is the suggestion that the spectrum below the
knee is dominated by a single nearby source and that the knee corresponds to
the maximal energy of this specific source~\cite{Erlykin:1997bs,Bouyahiaoui:2018lew}.

All these models lead to a rigidity-dependent sequence of knees at
$ZE_{\rm k}$, a behaviour first suggested by Peters~\cite{Peters61,Zatsepin62}.
Measurements of the nuclear composition of the CR flux can therefore not
distinguish between them. Alternative ways to extract information about the
CR knee are therefore needed. One possibility to infer the CR spectrum outside
the Solar neighbourhood is to use gamma-ray observations.
In this case, one uses the knowledge of the differential hadronic production
cross section of photons to infer the shape of the primary CR flux.
The first attempts to apply this method used relatively nearby giant molecular
clouds seen by Fermi-LAT~\cite{Neronov:2011wi,Kachelriess:2012fz,Dermer:2012bz,Neronov:2017lqd}.
Subsequently, gamma-ray observations were also used to derive the CR spectrum
as a function of Galactic longitude~\cite{Neronov:2015vua,Yang:2016jda,Aharonian:2018rob}. However, the relatively large errors of these early works prevented clear conclusions.

In this article, we aim to address the question of whether the CR knee observed locally is a 
global property of the Galactic CR spectrum or if it is a local feature, using new gamma-ray observations from LHAASO. As a first step, 
we derive in Section~\ref{sec:CR_model}
a model for the locally measured CR intensity and composition, taking into account, among others,
recent LHAASO measurements of the  mean logarithmic mass  $\langle\ln(A)\rangle$ 
in the knee region. Then, 
in Section~\ref{sec:gamma_model}, we work out predictions for the shape of the diffuse gamma-ray emission due to cosmic ray interactions in the interstellar medium, expected in this model. In Section~\ref{sec:results}, we compare these predictions with the data on diffuse emission from the Galactic disk and show that the predicted break in the spectrum of diffuse gamma-ray emission is not consistent with LHAASO measurements in the TeV--PeV energy range.
As a result, we conclude that the multi-messenger (cosmic-ray and gamma-ray) data suggest that the locally observed knee of the  CR spectrum is a local feature. We argue that a knee-like break 
is a generic property of the Galactic CR spectrum, but that its energy is a factor of order ten lower in the outer and inner Galactic disk than close to the Sun.

\section{Cosmic-ray model}
\label{sec:CR_model}

We aim to derive in this section a phenomenological model for the local CR intensity and composition 
fitting the recent CR data from different ground-based and satellite experiments. We assume that  different nuclei accelerated by a given source population 
follow a Peters' cycle, i.e., that their rigidity spectra are identical. As a functional form for the 
fit function for the contribution of the element $j$ with charge $Z$ from the $i$th population 
to the CR intensity, we use a power law with exponential cutoffs both at low and high energies,
\be
I_{i,j} = \frac{\mathcal{N}_{i,j}}{(E_{i0,j})^{2.5}} \left(\frac{E}{E_{i0,j}}\right)^{-\gamma_i}  \exp\left({-\frac{E_{i0,j}}{E}}-{\frac{E}{E_{i1,j}}}\right) .
\label{eq:power_law}
\ee
Here, $E$ is the total energy, $\mathcal{N}_{i,j}$ is a normalisation constant, $\gamma_i$ is the index of the power law and $E_{i0,j}$ ($E_{i1,j}$) 
are the low (high) energy cut-offs. Assuming Peters' cycles means that the cut-off energies  
for an element $j$ with charge $Z_j$ are shifted by a factor $Z_j$ compared to those of protons, e.g., $E_{i0,j} = Z_jE_{i0}$.
We add additional CR source populations until the fit to the data becomes acceptable.

For the CR intensities, we use data\footnote{Part of this data was gathered thanks to \cite{Maurin_2014, Maurin_2020, Maurin_2023}.} from AMS-02~\cite{AMS02_light,AMS02_Mg,AMS02_Fe}, DAMPE~\cite{DAMPE_protons,DAMPE_He,DAMPE_CNO}, CALET~\cite{CALET_protons, CALET_He, CALET_Heavy, CALET_Fe}, 
HAWC~\cite{HAWC_all}, IceCube/IceTop~\cite{data_Icecube} and iron data from the 
Pierre Auger Observatory (PAO)~\cite{Auger_fractions, Auger_all}.
In addition, we use the recent measurement of the all-particle intensity and the mean logarithmic mass $\langle\ln A\rangle$ performed by LHAASO~\cite{data_LHAASO:2024}.  In the case of LHAASO and the PAO, we choose the data obtained applying the EPOS-LHC hadronic interaction model~\cite{epos-lhc}.

In order to be consistent with the common practice of indirect CR experiments (such as the PAO), we divide the CR all-particle spectrum for all experiments into four groups: protons, helium, intermediate, and heavy elemental groups.
From the intensity of a given element with charge $Z$, 
the relative contribution of this element to the intermediate group is computed as
\be
f_{Z} =  \frac{ Z_{\rm Fe} - Z}{ Z_{\rm Fe} - Z_{\rm CNO} } 
\ee
with $Z_{\rm CNO} = 7$ and $Z_{\rm Fe} = 26$. The remainder, $(1 - f_{Z})$, is put into the heavy group, so that the all-particle intensity is also the sum of our four group intensities.  

Because of the systematic uncertainties in the energy calibration of various experiments, we rescaled the
energy scale of each experiment. Since AMS-02 has relatively small errors and a rather complete data set
covering most elements~\cite{AMS02_light}, we used it as reference. The data of other experiments are  
rescaled such that the  all-particle spectrum is a continuous function of energy. 
Since AMS-02, DAMPE and CALET have not published  all-particle spectra, we proceed in their case as follows. 
For AMS-02, we compute the all-particle spectrum summing up all elements measured. 
In the case of CALET, we rescale the energy of each elemental group individually to match the measurements of AMS-02.
DAMPE has not published results for elements heavier than oxygen. Therefore we use the proton data to
derive the overall scale factor, and apply it to the other elemental groups. The intensity of the intermediate group is then directly scaled to the one measured by AMS-02.
All the derived scaling factors are well within the energy measurement uncertainties, and are listed in Table~\ref{tab1}.

Performing a $\chi^2$ fit, the uncertainties are computed as the quadrature sum of the statistical and
systematic uncertainties. In the case of non-symmetric uncertainties, upper and lower uncertainties 
are computed in the same way, choosing the appropriate one depending on whether the fit point is above or 
below the data point. The fit is performed using the {\tt lmfit} library~\cite{lmfit}, including the intensities of the
elemental CR groups given by AMS-02, DAMPE/CALET, and the intensity of the iron group as measured by the PAO. Around the knee energy, we use LHAASO data on the all-particle spectrum and $\langle\ln A\rangle$.
The values of the parameters obtained through the fit are presented in the appendix in Tables~\ref{tab2} and \ref{tab3}.

\begin{figure*}
   \begin{subfigure}[t]{1.99\columnwidth}
        \includegraphics[width = 0.99\columnwidth]{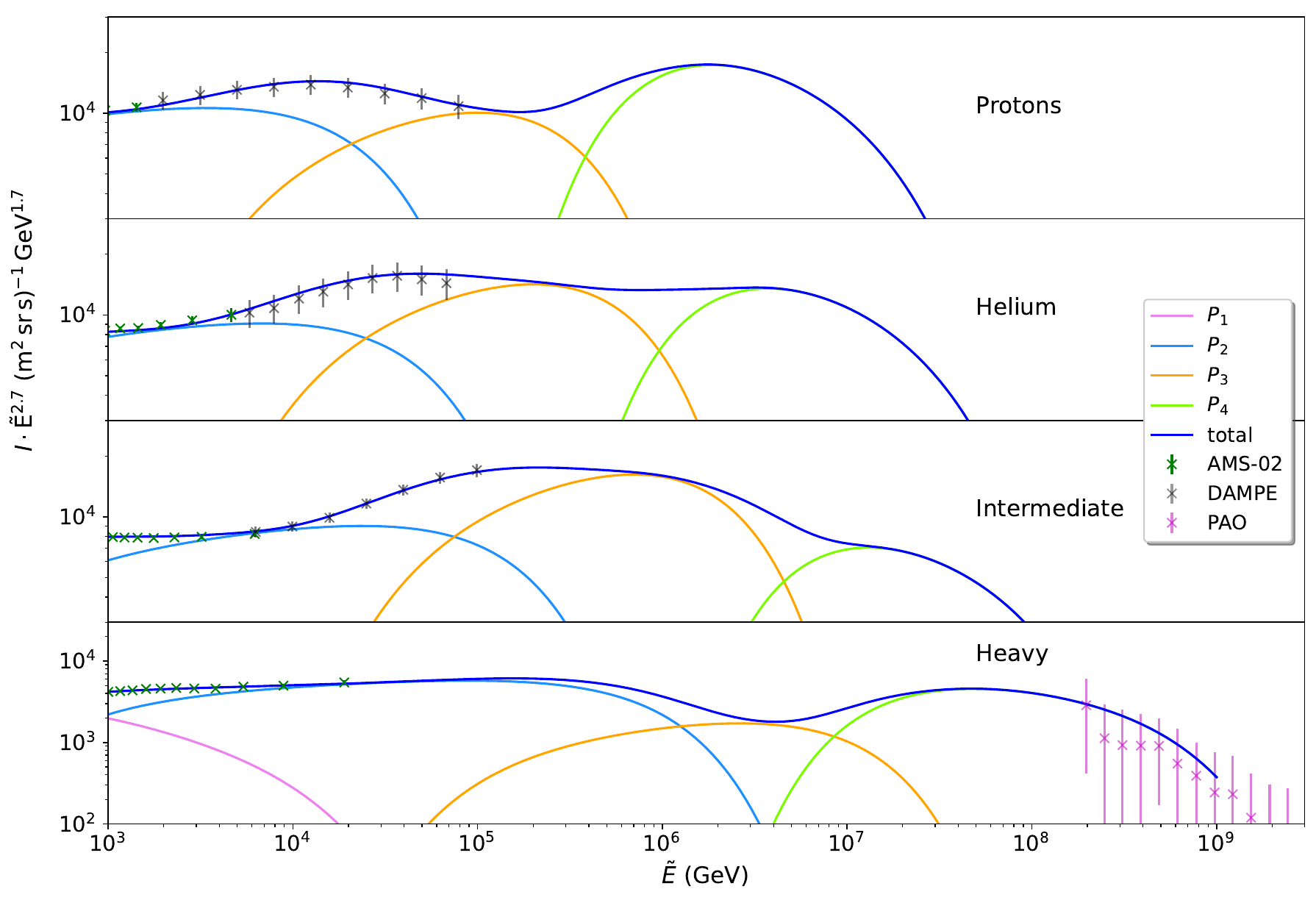}
    \end{subfigure}
    \hfill
    \begin{subfigure}[b]{1.99\columnwidth}    
        \includegraphics[width = 0.99\columnwidth]{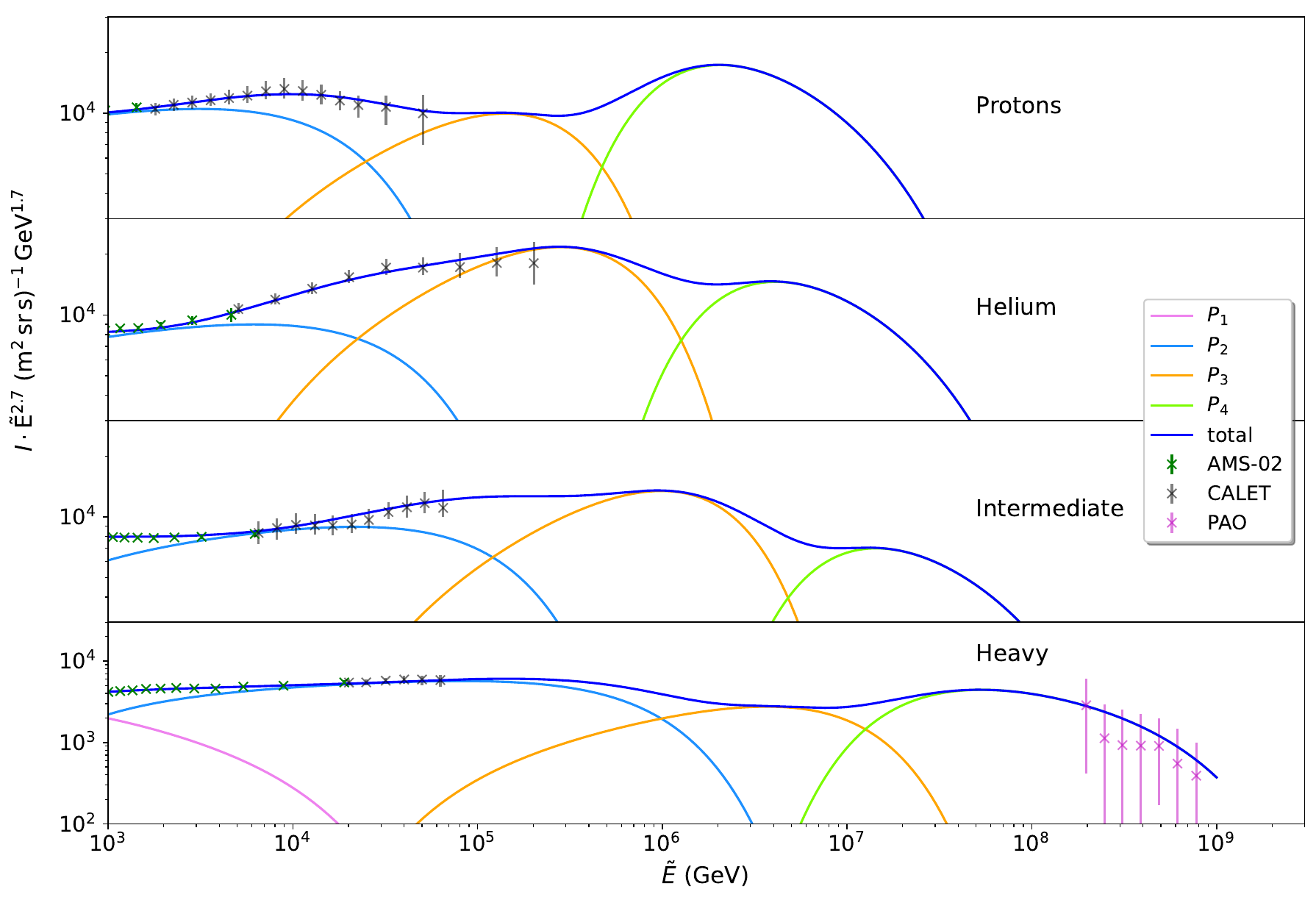}
    \end{subfigure}
    \caption{The fits of the CR intensity of four elemental groups as a function of the rescaled energy compared to data from AMS-02~\cite{AMS02_light, AMS02_Mg, AMS02_Fe}, DAMPE~\cite{DAMPE_protons, DAMPE_He, DAMPE_CNO}, 
    CALET~\cite{CALET_protons, CALET_He, CALET_Heavy, CALET_Fe}, LHAASO~\cite{data_LHAASO:2024} and PAO~\cite{Auger_fractions, Auger_all}; 
    in the top (bottom) figure, the fit uses DAMPE (CALET).   
    }
    \label{fits}
\end{figure*}

\begin{figure*}
    \includegraphics[width = 1.99\columnwidth]{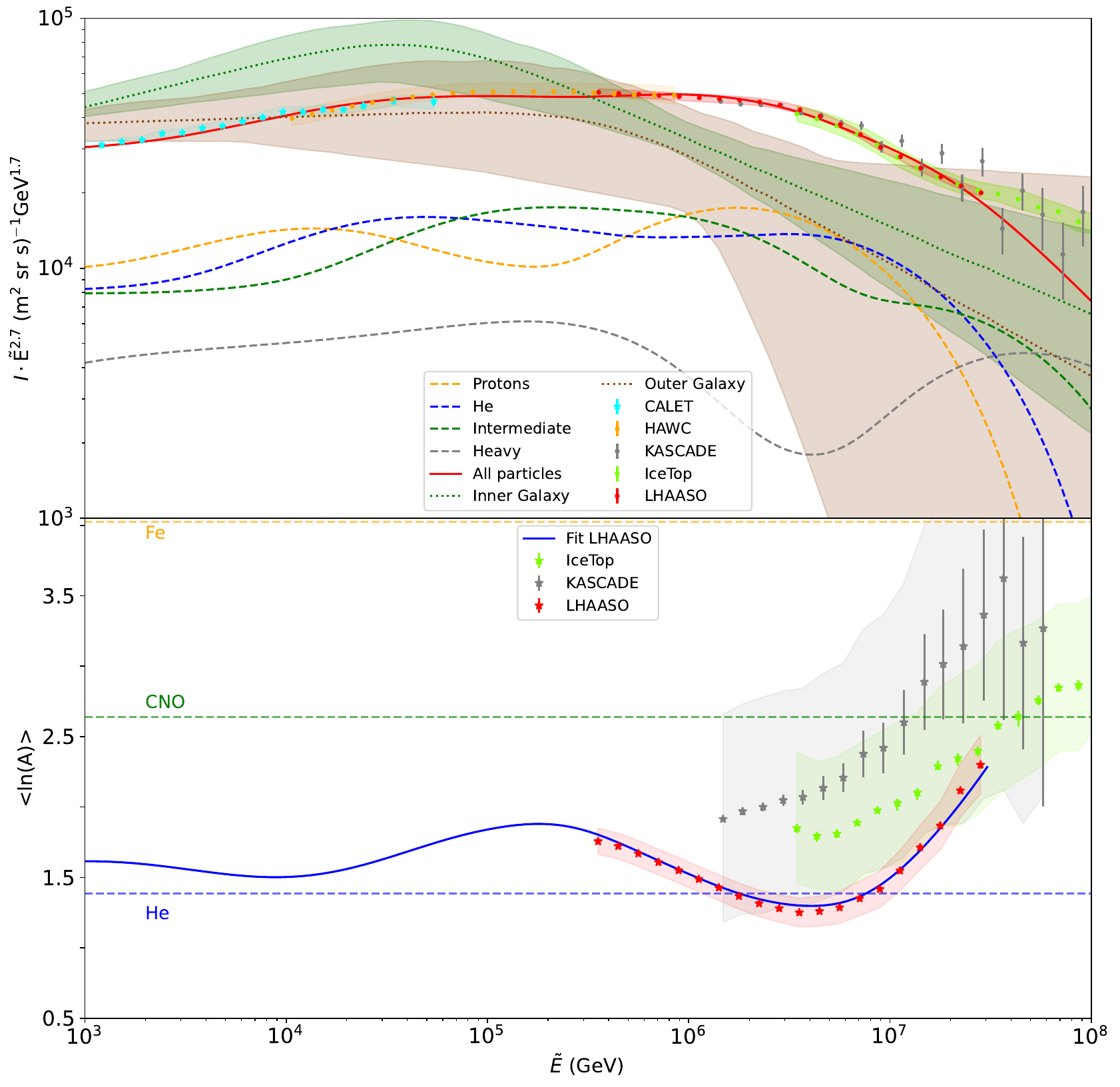}
    \caption{Top: All-particle spectrum (solid line) and partial contributions (dahed lines) of the fitted CR spectrum with sata is from CALET~\cite{CALET_protons, CALET_He, CALET_Heavy, CALET_Fe}, HAWC~\cite{HAWC_all}, KASCADE~\cite{finger}, IceTop~\cite{data_Icecube} and LHAASO~\cite{data_LHAASO:2024} together with the all-particle spectrum (dotted lines) and their $2 \sigma$ uncertainty band from a fit to the gamma-ray data. Bottom: Mean logarithmic mass from our model as a function of energy compared to data from IceTop~\cite{data_Icecube}, KASCADE~\cite{finger} and LHAASO~\cite{data_LHAASO:2024}.}  
\label{fig:all_lnA_DAMPE_LHAASO}
\end{figure*}

Since the data of  DAMPE and CALET differ slightly, we perform
two separate fits including only one of these two experiments each time.
The resulting CR spectra  are shown in the top and bottom panels of  Fig.~\ref{fits}
for DAMPE and CALET. In each panel, we show the
fluxes of the four elemental groups. 
The resulting total intensity and mean logarithmic mass $\langle\ln(A)\rangle$ are shown in 
Fig.~\ref{fig:all_lnA_DAMPE_LHAASO}. For both cases, we find an excellent  agreement with the  
$\langle\ln(A)\rangle$ data of LHAASO. In addition, we show $\langle\ln(A)\rangle$ from 
IceTop~\cite{data_Icecube} and KASCADE~\cite{finger}. While the IceTop data agree with 
LHAASO given their large systematic uncertainty, there is a tension between the earlier
KASCADE data and LHAASO. As a result, the inferred composition of CRs in the knee region
based on LHAASO data differs from earlier analyses.

In total, four source populations are required to obtain a good description
of the data. Note that the iron component of the fourth Galactic population, which constitutes
the end of the Galactic CR spectrum, describes the iron flux measured
by the PAO up to $\simeq 10^{18}$\,eV. 
We also tried to fit the data with only three components, using e.g.\ a broken power law for the component around the knee instead of two Peters' cycles. However, 
in that case,  the fits were not satisfactory: while the all-particle spectrum from LHAASO could 
be fitted, the last points of DAMPE and LHAASO’s measurements of the mean logarithmic mass could not be
reproduced. It is the additional dip seen in Fig.~\ref{fig:all_lnA_DAMPE_LHAASO} in $\langle\ln(A)\rangle$
which requires the introduction of a fourth population of CR sources. The reduced $\chi^2$ with three populations
was around 1.4, while with four components it diminished to $0.38$.


\section{Connecting CR and photon fluxes}
\label{sec:gamma_model}

Cosmic ray protons and nuclei interact in the interstellar medium, producing neutral pions that subsequently decay into gamma rays. This leads to the diffuse "glow" of the interstellar medium in gamma rays. The diffuse gamma-ray emission from the Milky Way disk is the dominant gamma-ray source on the sky in GeV-TeV energy range \cite{2012ApJ...750....3A}. The diffuse emission from the Galactic disk has now been measured up to PeV energy by Tibet AS/$\gamma$~\cite{2021PhRvL.126n1101A} and LHAASO~\cite{LHAASO:2023gne}.

In the first approximation, the spectrum of gamma-ray emission from the decays of neutral pions produced by a power-law CR proton or nuclei population with a slope $\alpha_A$ is also a power law with the slope $\alpha_\gamma$ close to the slope of the parent proton spectrum, while the gamma-ray emissivity depends on the atomic number $A$ and the slope $\alpha$ though the so-called $Z$-factor, or nuclear enhancement factor~\cite{Gaisser:1990vg}. In general, however, neither the use of $Z$-factors
or of a nuclear enhancement factor to account for heavier nuclei is
justified~\cite{Kachelriess:2014mga}. For a more detailed calculation of the gamma-ray intensity as a function of the sky direction, $(l,b)$ and photon energy $E$ we use the {\tt AAfrag} package~\cite{Kachelriess:2019ifk,Koldobskiy:2021nld} to evaluate the production cross-sections that enter the expression for the overall gamma-ray flux
\begin{align}
  I_\gamma(E,l,b) & = \frac{c}{4\pi}\sum_{A,A'}  \int_0^\infty \!\!\d s\,
  n_{\rm gas}^A(x) \e^{-\tau(s,E)}\nn\\
 &  \int_E^\infty \!\!\d E' \: \frac{\d\sigma^{A'A\to X\gamma}(E',E)}{\d E}
  \frac{d N^{A'}_{\rm CR}}{\d V\,\d E'},
\end{align}
where $x=\{s,l,b\}$ is the 3-dimensional spherical coordinate system ($s$ is the distance along the line of sight in the $(l,b)$ direction in Galactic coordinates) and $\d\sigma^{A'A\to X\gamma}(E',E)/\d E$ is the differential
cross section to produce photons with energy $E$ in the scattering of a primary
nucleus $A'$ with energy $E'$ on target nuclei $A$ with density $n_{\rm gas}^A$.  
The optical depth $\tau$ takes into account the absorption of photons
due to pair production on cosmic microwave background (CMB) photons. 
We neglect the absorption on the extragalactic background light and the starlight in the Milky Way,
which add only a minor correction relative to the absorption on the CMB~\cite{Vernetto2016}. 

We assume that the dependence of the spectral CR density on  energy $E$ and position $x$ factorises,
\be
I_\mathrm{CR}=\frac{c}{4 \pi}\frac{d N^{A'}_{\rm CR}}{\d V\,\d {E}} = \frac{c}{4 \pi}n_{\rm CR}^{A'}(x)
 \frac{d N^{A'}_{\rm CR}}{\d {E}},
\ee
leading to the factorisation of the photon intensity,
\begin{align}
  I_\gamma(E,l,b) & = \frac{c}{4\pi}\sum_{A,A'}  F^{A'A}(E,l,b) G^{A'A}(E)
\end{align}
with
\be \label{eq:f}
F^{A'A}(E,l,b)= \int_0^\infty \!\!\d s\, n_{\rm gas}^A(x)n_{\rm CR}^{A'}(x) \e^{-\tau(s,E)}
\ee
and
\be
 G^{A'A}(E) = 
 \int_E^\infty \!\!\d E' \: \frac{\d\sigma^{A'A\to X\gamma}(E',E)}{\d E}
  \frac{d N^{A'}_{\rm CR}}{\d E'} .
\ee
The intensity of neutrinos produced in hadronic interactions is calculated in the 
same way, setting $\tau=0$ in Eq.~(\ref{eq:f}).

For the composition of the gas in the interstellar medium, we use 0.91:0.09
as the mass ratio of hydrogen and helium, i.e., we neglect heavier elements. 
For the spatial distributions $n_{\rm gas}^A(x)$ and $n_{\rm CR}^{A'}(x)$, 
we employ the model of \citet{Lipari2018}.

\section{Results}
\label{sec:results}

The LHAASO collaboration has reported an analysis of the diffuse $\gamma$-ray emission  from 
the Galactic plane up to PeV energies~\cite{LHAASO:2023gne}, for two representative regions of 
the Galactic disk: the inner Galaxy ($25^\circ < l < 125^\circ$, $|b|<5^\circ$)
and the outer Galaxy ($125^\circ < l < 235^\circ$, $|b|<5^\circ$).
The $\gamma$-ray flux measured with  Fermi-LAT data for the same two regions of sky was calculated in Ref.~\cite{Zhang:2023ajh}.  

\begin{figure*}
    \includegraphics[width = 1.99\columnwidth]{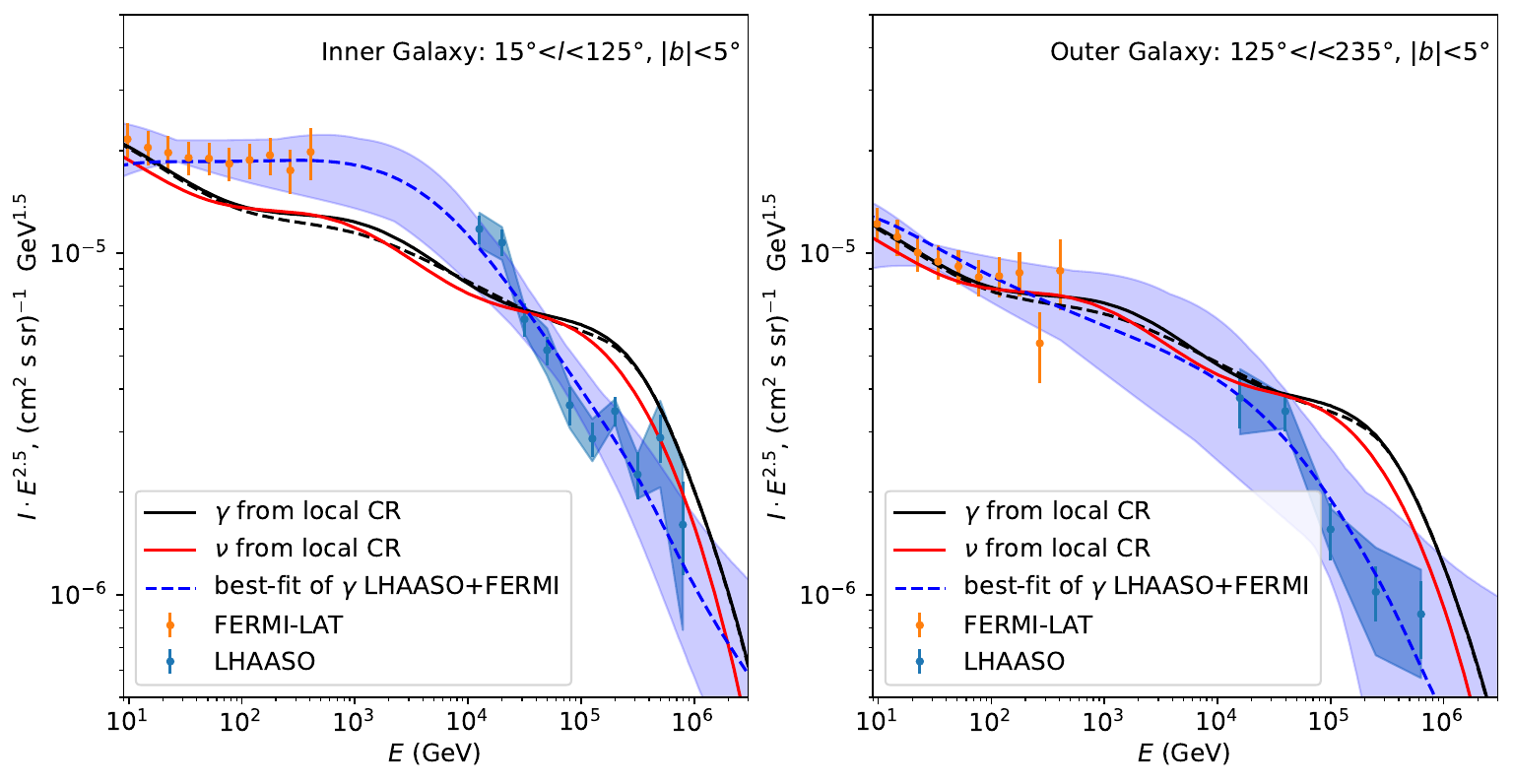}
    \caption{Gamma-ray spectrum in the LHAASO inner and outer Galaxy -- measurements by Fermi-LAT and LHAASO and expected gamma-ray flux considering the CR flux to be the same as in the Earth's vicinity.
    The solid black line corresponds to DAMPE+LHAASO model, while the dashed line is for CALET+LHAASO.
    The expected neutrino flux in the LHAASO Inner and Outer Galaxy regions obtained with the DAMPE+LHAASO CR model is shown by a red line; 
    the best-fit of Fermi-LAT+LHAASO gamma-ray data by a blue dashed line with its 95\% C.L. indicated with blue shading.}
    \label{fig:results}
\end{figure*}

These LHAASO and Fermi-LAT measurements are compared with the predictions of the models described in the previous sections in Fig.~\ref{fig:results},
where the spectrum of diffuse gamma-ray emission from hadronic interactions for the aforementioned regions is shown.
This spectrum is calculated summing the spectra of gamma rays produced by interactions of all four elemental groups contributing to the CR spectrum (modeled in Section~\ref{sec:CR_model}), 
using the methodology described in Section~\ref{sec:gamma_model}. 
We also produced a prediction for the expected all-flavour neutrino flux using the same methodology as described above for the gamma emission (neglecting the absorption).

The left panel of Fig.~\ref{fig:results} shows a comparison of the predicted gamma-ray spectrum with the diffuse gamma-ray flux measurements in the inner Galaxy.
We note that for our prediction of the photon flux,
we have applied the same sky mask as in Refs.~\cite{Zhang:2023ajh,LHAASO:2023gne}. One can notice that the model under-predicts the diffuse flux at the high-energy end of the Fermi-LAT energy range. This can be due to a range of reasons, such as variations of the slope of the CR spectrum across the Galactic disk \cite{Neronov:2015vua,Yang:2016jda} or due to the existence of an unresolved source population, such as pulsar halos~\cite{Yan:2023hpt,Dekker:2023six}. The model also over-predicts the diffuse flux in the LHAASO energy range above 100\,TeV. This inconsistency indicates that the CR spectrum in the inner Galaxy cannot have a knee at the same energy as the knee of the local CR spectrum. The inconsistency can be 
removed by lowering the energy of the knee, for example, by 
reducing in the inner Galaxy the maximal energy of
the 4th Peters' cycle. 

The same inconsistency of the model predictions with LHAASO data can be seen in the right panel of Fig.~\ref{fig:results} which shows the spectrum of diffuse emission from the outer Galaxy. One can see that, contrary to the inner Galaxy case, the model reproduces well the Fermi-LAT diffuse emission spectrum up to the TeV energy range. Moreover, the model is also consistent with the first two data points of LHAASO in the 10\,--\,100\,TeV energy range. However, the model predictions in the $E>100$\,TeV band are largely above the LHAASO measurements. This means that the knee of the CR spectrum should be at a lower energy also in the outer Galaxy. 

\begin{figure}
    \centering
    \includegraphics[width=1\linewidth]{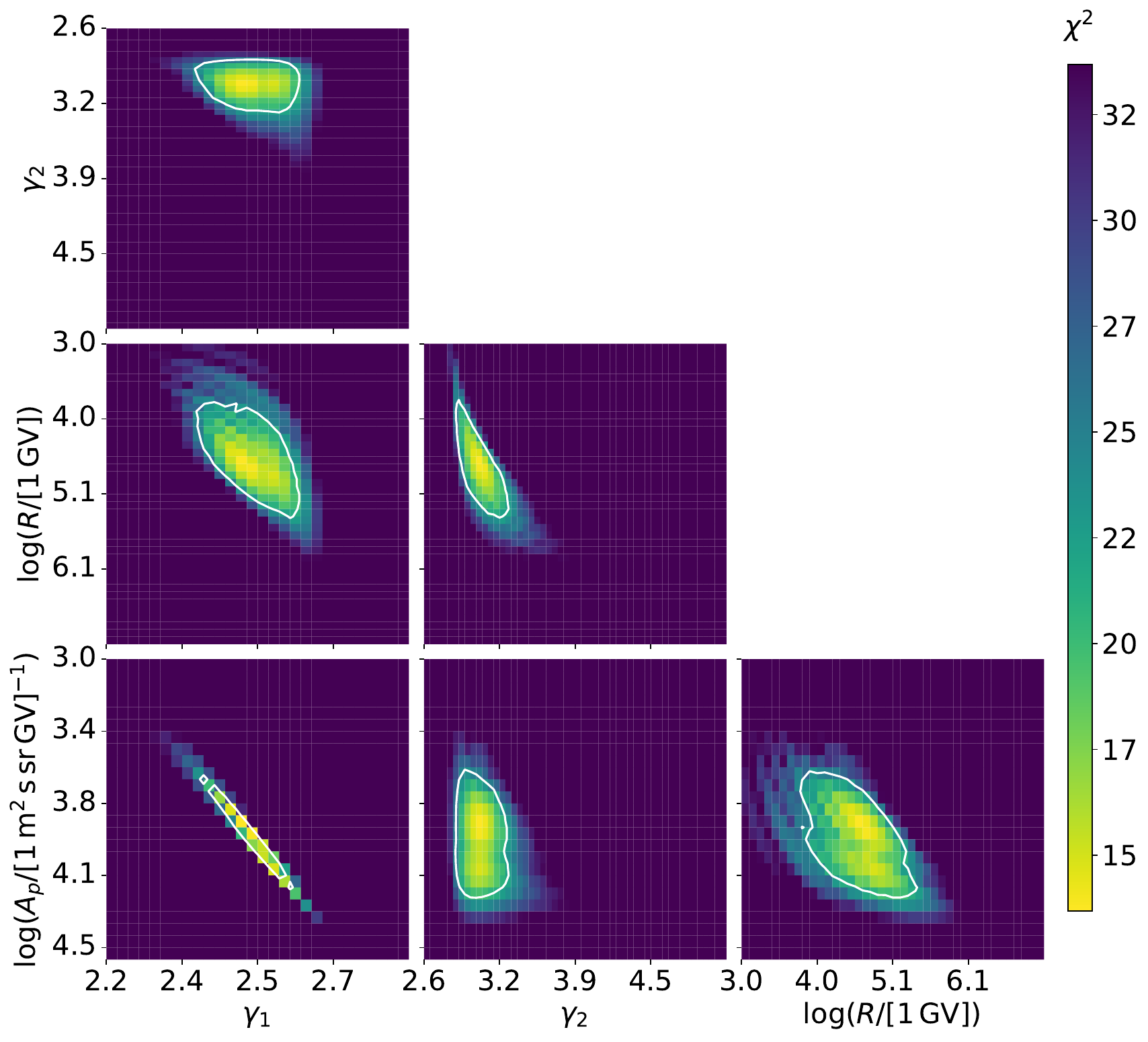}
    \caption{Parameter space for the broken-power law fit of the combined LHAASO+Fermi-LAT data for the inner Galaxy region; white contours denote the 95\% C.L. }
    \label{fig:gamma_inner_galaxy_params}
\end{figure}

We have also made a prediction of the expected CR intensity based on the diffuse gamma-ray observations.
In order to fit the LHAASO data, we assume again that the knee-like structure
in the CR energy spectra follows a Peters' cycle, i.e., that the 
rigidity spectra of different CR nuclei is the same, differing only by their normalisation,
\be
 \frac{\d N^{A'}_{\rm CR}}{\d E'} =  N_0^A R^{\gamma_1}
 \left[ 1+ \left(\frac{R}{R_b}\right)^s \right]^{\frac{\gamma_2-\gamma_1}{s}} \frac{\d R}{\d E} .
\ee
Here, $\gamma_{1/2}$ are the asymptotic slopes and $s$ describes the
sharpness of the transition. For our calculations, we have used $s=2$. Since the inversion of the integral equation~(5) is not unique, we have to constrain the CR spectra by hand:  We include only the two most important nuclei, protons and helium, fixing moreover their relative normalisation as $N_0^{\rm He}=N_0^p/4$.
Then we have used an adaptive grid in the parameter space $\{N_0^p,\gamma_1,\gamma_2,R_b\}$, calculating for every parameter set $G^{A',A}(E)$ and then $\bar I_\gamma(E)$ for both the inner and outer Galaxy regions. For each realisation of the fit, we have calculated the $\chi^2$ value of the modeled gamma-ray spectrum compared to the observational data of LHAASO and Fermi-LAT. The fits were performed independently for the inner and outer Galaxy regions, the best-fit is shown in Fig.~\ref{fig:results} as a blue dashed line. In the case of the inner Galaxy, a knee-like break around 20\,TeV is clearly visible.
Using Monte-Carlo modeling, we have also assessed the uncertainties of the fit parameters. 
The resulting 95\% C.L. band of the gamma-ray flux is shown as a blue band, while the uncertainties of the fit parameters are shown in Fig.~\ref{fig:gamma_inner_galaxy_params}.
In particular, we have determined the location of the CR knee for the inner Galaxy region to be in range $35^{+194}_{-28}$\,TV within 95\% C.L.\ (defined as parameter space under the condition of $\chi^2<\chi_\mathrm{min}^2+9.49$~\cite{Press2007}). Therefore the energy of the CR knee appears to be significantly lower in the inner Galaxy region than in vicinity of the Solar system. We obtain similar results for the outer Galaxy region, however, their significance is lower.

The all-particle CR spectrum obtained in the fit to the gamma-ray data in the inner and outer Galaxy is also shown in Fig.~2
as dotted lines together with the corresponding $2\sigma$ uncertainty band. In both cases, a steepening at an energy clearly below 4\,PeV is visible
in the best-fit line. In the inner Galaxy, where the uncertainties of the fit are smaller, the break is more pronounced and required within
all variations allowed by the $2\sigma$ uncertainty band.

Overall, one can notice that the knee of the CR spectrum introduces a knee-like feature in the gamma-ray model spectrum. To understand if the knee is a generic feature of the Galactic CR spectrum, one needs to identify the gamma-ray knee feature in the spectrum of the diffuse emission from the Galactic disk. It is clear that the currently available data are inconsistent with a gamma-ray knee at $\sim 300$\,TeV  as expected in the model of a "universal knee" at a fixed energy for the entire Galactic CR population. Instead, a combination of the Fermi-LAT and LHAASO measurements, shown in the left and right panels of 
Fig.~\ref{fig:results} suggests that the spectrum of diffuse gamma-ray emission may actually have a knee feature at $E_{k,\gamma}\sim 30$\,TeV energy, both in the inner and the outer Galaxy. This may be an indication of the existence of the knee in the average Galactic cosmic-ray spectrum. However, the energy of this knee should be $E_{\rm k}^{\rm GalCR}\sim 10E_{\rm k}^\gamma\sim 300$\,TeV, a factor of ten lower than the energy of the locally observed knee.

\section{Conclusion}

The multi-messenger CR and gamma-ray analysis presented in this paper suggests that the knee of the cosmic ray spectrum at $E_k\simeq 4$~PeV energy is a feature of the population of CRs in the neighbourhood of the Solar system, rather than a universal property of the overall Galactic CR population. 
A potential explanation for this difference is that the local CR spectrum is influenced
by a rather young, local source with high maximal energy.
Both the $\gamma$-ray data from the inner and outer Galaxy show a break at an energy which is an order-of-magnitude lower than inferred from the local CR knee. 
This suggests that the overall Galactic CR knee is at an energy an order of magnitude lower than the local one.
This hypothesis can be verified by a clear identification of the knee-like feature in the spectrum of diffuse gamma-ray emission from the Galactic plane  and the verification of the pion-decay nature of the gamma-ray signal of this feature. Such measurements are possible with a combination of gamma-ray (LHAASO, CTAO) and neutrino (KM3NET, IceCube) measurements in the near future.


\appendix
\section{Fit parameters}

We present in this appendix the parameters of our CR model. The factors $\tilde{E} = E/f$ used to rescale the energy scale
of the various experiments are collected in Table~\ref{tab1}, where we also show the energy resolution of those experiments. As can be seen, the energy measured by DAMPE is rescaled by the same factor for all groups: because we lacked most elements from the CNO group, we instead directly adjusted the intensity of this group to AMS-02, multiplying it by $1.575$. Lacking only two elements  for CALET compared to AMS-02 (N and Al), we rescaled the energy for the intermediate and heavy groups directly on the corresponding AMS-02 measurements. Concerning the energy resolution of the different experiments, we generally provided lower bounds, either because this resolution is energy dependent, in which case we took the best resolution, or because authors only provided a rough estimate of this quantity, in which case we also took the most constraining resolution. One can notice that the rescalings we proposed remain well below the energy resolution provided by the experiments.

Concerning the fits parameters (see Eq.~\ref{eq:power_law}), the energy thresholds $E_{ij}$ and the slopes $\gamma_i$ are  presented in Table~\ref{tab2},
while the normalisation constants $N_i$ are shown in Table~\ref{tab3}. 
The models are composed of four populations, each of which is described by seven parameters: two energy thresholds, the index of the power law, and four normalisations parameters, one for each elemental group. Tables~\ref{tab2} and ~\ref{tab3} show parameters obtained for our two models, the first one fitting data from AMS-02, DAMPE, LHAASO and the PAO, while the second one fits CALET instead of DAMPE.

\begin{table*}[]
\centering
\begin{tabular}{|>{\centering\arraybackslash}m{1.8cm}||>{\centering\arraybackslash}m{1cm}>
{\centering\arraybackslash}m{1.2cm}>{\centering\arraybackslash}m{1.2cm}>{\centering\arraybackslash}m{1.2cm}>{\centering\arraybackslash}m{1.3cm}>{\centering\arraybackslash}m{1.3cm}>{\centering\arraybackslash}m{1.3cm}>{\centering\arraybackslash}m{1.4cm}>
{\centering\arraybackslash}m{1.4cm}|}
     \hline
      Experiments & AMS-02 & CALET H\&He & CALET Inter. & CALET Heavy  & DAMPE & HAWC & LHAASO &  IceTop & PAO   \\
     \hline
      Factor $f$ & 1 & 0.99 & 0.87 & 0.90 & 1.0 & 1.05 & 1.0 & 1.03 & 0.91 \\
      Energy resolution &   & $30\%$ & $30\%$ & $30\%$ & $10\%$ &  $7\%$ & $5.5\%$ & $9\%$ & $\sim 14\%$ \\
      \hline

\end{tabular}
\caption{Factors used to rescale the experiments as $\tilde{E} = E/f$.}
\label{tab1}
\end{table*}

\begin{table*}[]
\centering
\begin{tabular}{|p{1.8cm}||p{1cm}p{1cm}p{1cm}p{1cm}p{1cm}p{1cm}p{1cm}p{1cm}p{1cm}p{1cm}p{1cm}|}
     \hline
      Parameters & $ E_{11} $ & $E_{20}$ & $E_{21}$ & $E_{30}$ & $E_{31}$ & $E_{40}$ & $E_{41}$ & $\gamma_1$ & $\gamma_2$ & $\gamma_3$ & $\gamma_4$ \\
     \hline
      DAMPE  & 4.06e2 & 2.20e1 & 2.81e4 & 4.41e3 & 3.08e5 & 7.90e5 & 1.90e7 & 3.25 & 2.59 & 2.42 & 3.02 \\
      CALET  & 4.06e2 & 2.20e1 & 2.58e4 & 2.95e3 & 2.67e5 & 1.20e6 & 2.30e7 & 3.25 & 2.59 & 2.19 & 3.19 \\
      \hline
\end{tabular}
\caption{Energy thresholds $E_{i0},E_{i1}$ (in GeV) and exponents $\gamma_i$  of the power laws used for the description of the CR energy spectra (Eq.~\ref{eq:power_law}).}
\label{tab2}
\end{table*}

\begin{table}[]
\centering
\begin{tabular}{|p{3.5cm}||p{1cm}p{1cm}p{1cm}p{1cm}|}
     \hline
     Parameters & $\mathcal{N}_1$ & $\mathcal{N}_2$ & $\mathcal{N}_3$ & $\mathcal{N}_4$ \\
     \hline
      DAMPE - Protons & 2.03e4 & 3.67e3 & 1.13e3 & 2.55e3  \\
      DAMPE - Helium & 1.06e4 & 2.73e3 & 1.38e3 & 1.73e3 \\
      DAMPE - Intermediate & 7.20e3 & 2.10e3 & 1.23e3 & 7.01e2 \\
      DAMPE - Heavy & 2.50e3 & 1.03e3 & 1.00e2 & 3.50e2 \\
      \hline
      CALET - Protons & 2.03e4 & 3.67e3 & 4.86e2 & 2.70e3 \\
      CALET - Helium & 1.06e4 & 2.73e3 & 9.21e2 & 1.98e3 \\
      CALET - Intermediate & 7.20e3 & 2.10e3 & 4.43e2 & 7.35e2 \\
      CALET - Heavy & 2.50e3 & 1.03e3 & 7.00e1 & 3.60e2 \\
      \hline
\end{tabular}
\caption{Values of the normalisation constants $N_i$ in $(\mathrm{m\,s\,sr})^{-1}\mathrm{GeV^{1.5}}$ (Eq.~\ref{eq:power_law}). 
}
\label{tab3}
\end{table}


\end{document}